# Omnidirectional nonreciprocal absorber realized by the magneto-optical hypercrystal

Shengyu Hu[1], Juan Song[1], Zhiwei Guo[1,2,*], Yong Sun[1], Yunhui Li[1], Haitao Jiang[1,*], Fusheng Deng[2], Lijuan Dong[2], and Hong Chen[1]

[1]Key Laboratory of Advanced Micro-structure Materials, MOE, School of Physics Science and Engineering, Tongji University, Shanghai 200092, China

[2] Shanxi Provincial Key Laboratory of Microstructure Electromagnetic Functional Materials, Shanxi Datong University, Datong, 037009, China

**Abstract:** Photonic bandgap design is one of the most basic ways to effectively control the interaction between light and matter. However, the traditional photonic bandgap is always dispersive (blueshift with the increase of the incident angle), which is disadvantageous to the construction of wide-angle optical devices. Hypercrystal, that the photonic crystal with layered hyperbolic metamaterials (HMMs), can strongly modify the bandgap properties based on the anomalous wavevector dispersion of the HMM. Here, based on phase variation compensation between HMM and isotropic dielectric layers, we propose for the first time to design nonreciprocal and flexible photonic bandgaps using magneto-optical HMMs in one-dimensional photonic crystals. Especially for the forward and backward incident light, the blueshift and dispersionless of the forward and backward cavity modes are designed respectively to realize the interesting omnidirectional nonreciprocal absorber. Our results show high (low) absorption about 0.99 (0.25) in an angle range of 20-75 degrees for the forward (backward) incident light at the wavelength of 367 nm. The nonreciprocal omnidirectional cavity mode not only facilitates the design of perfect unidirectional optical absorbers working in a wide-angle range, but also possesses significant applications for all-angle reflectors and filters.

# I. INTRODUCTION

Band engineering plays an important role in the control of the light, and is of great significance in fundamental and applied physics. For the photonic crystals (PCs) with multiple scattering mechanism, the bandgap has been widely studied for suppressing the spontaneous emission and localization of photons [1, 2]. Especially, one-dimensional (1D) PCs composed of altering dielectric layers with different materials have attracted people's great attention because of the simple structure and important applications. However, it is well known that with the increase of the incident angle, the bandgap of all-dielectric 1D PC always changes to short wavelengths (blueshift). This property is come from the circle iso-frequency contour (IFC) of the dielectric. According to the Bragg condition, the bandgap of all-dielectric PC is formed when the sum of the propagating phases in the two dielectrics in one period is the integer times of $\pi$. Specially, the larger incident angle is, the smaller wavevector of light in the propagating direction is. As a result, when the incident angle increase, a larger wavelength is needed to obtain the fixed propagating phase to realize the destructive interference, thus leading to the blueshift of bandgap. This dispersive property of bandgap is not conducive for designing broadband optical devices, such as the all-angle absorbers, reflectors and filters [3].

Metamaterials, artificial materials composed of subwavelength unit cells, provide a powerful platform for manipulating the propagation of light [4-7]. Especially, 1D PCs with metamaterials, including double-negative metamaterials or two types of single-negative metamaterials, can manipulate the bandgap more flexibly and realize the interesting dispersionless bandgaps based on the mechanism of phase cancellation [8-12]. As one of the most unusual classes of metamaterials, hyperbolic metamaterials (HMMs) recently have attracted immense interest in a wide range of subject areas in physics and engineering [13-16]. Because of the abnormal dispersion property of HMMs, flexible control of the propagation of electromagnetic waves is realized, such as enhanced photonic density of states [17], anomalous refraction [18, 19], super-resolution imaging [20-22], cavities with anomalous scaling laws [23, 24], subwavelength waveguides [25, 26], and long-range energy transfer [27, 28]. In 2014, E. Narimanov theoretically proposed the hypercrystal, which combines the properties of PC with metamaterials [29]. Hypercrystal provides a new way to control the interaction between light and the matter [29, 30]. In particular, Xue *et al*. discovered that dispersionless gaps can be realized in 1D hypercrystal composed of layered HMMs and dielectrics [31]. The underlying physical mechanism comes from the phase-variation compensation effect in hypercrystal. The special IFC of HMM determines that the change of the propagating phase with the incident angle in HMM is opposite to that in the dielectric. Therefore, once the variation of propagating phase in the dielectric layers is compensated by the HMM layers, the bandgap will not change with the incident angle, thus forming the dispersionless bandgap [31-33]. This controlled bandgap in hypercrystal can be used to design some wide-angle devices, such as the sensors [34] and splitters [35].

Recently, magnetized metamaterials enable the exploration of new regime about the magneto-optical (MO) effect, including the enhanced nonreciprocal transmission and one-way surface waves [36, 37]. A natural question is whether MO HMM can be used to design hypercrystal with arbitrary control of band dispersion? In this work, based on the previous works mentioned above, we first propose the nonreciprocal and flexibly controlled photonic bandgaps in hypercrystal with MO HMMs. For the effective MO HMMs composed of subwavelength dielectric/metal/dielectric multilayers, the corresponding IFCs for the forward and backward incident waves can be flexibly tuned by the external magnetic field [37]. For example, the IFCs of forward and backward incident waves in MO HMMs can be designed as closed circle and open hyperbolic curves, respectively. According to the phase variation compensation between MO HMM and isotropic dielectric layers, nonreciprocal and flexible photonic bandgaps can be realized. Especially for the forward and backward incident light, the dispersionless and blueshift of the forward and backward cavity modes are designed respectively to realize the interesting omnidirectional nonreciprocal absorber. Our results provide a way to design the novel optical nonreciprocal devices with excellent performance by using the HMMs and would be very useful in various applications including optical isolators, circulators, sensors, reflectors, filters and switches.

This work is organized as follows: Sec. II covers the design of the design of MO HMM by the subwavelength dielectric/metal/dielectric stacks. Specially, the flexible control of band structure of the 1D MO hypercrystal with incident angle is studied; in Sec. III, the cavity structure based on the MO hypercrystal is carried out to realize the omnidirectional nonreciprocal absorption. Finally, Sec. IV summarizes the conclusions of this work

## II. BAND ENGINEERING BASED ON THE HYPERCRYSTAL CONSISTS OF MO HMM AND DIELECTRIC

As shown in Fig. 1, the MO hypercrystal proposed here is a 1D PC containing MO HMMs, which is denoted by $(AB)_N$. The MO HMM (layer A) is mimicked by subwavelength dielectric/MO metal/dielectric stacks as $(CDE)_M$. $N=10$ and $M=5$ denote the period numbers of PC and HMM, respectively. When the thickness of unit cell $d = d_C + d_D + d_E$ is far less than the wavelength of electromagnetic wave in the structure, the structure $(CDE)_M$ can be equivalent to MO HMM based on the effective medium theory (EMT) [37].

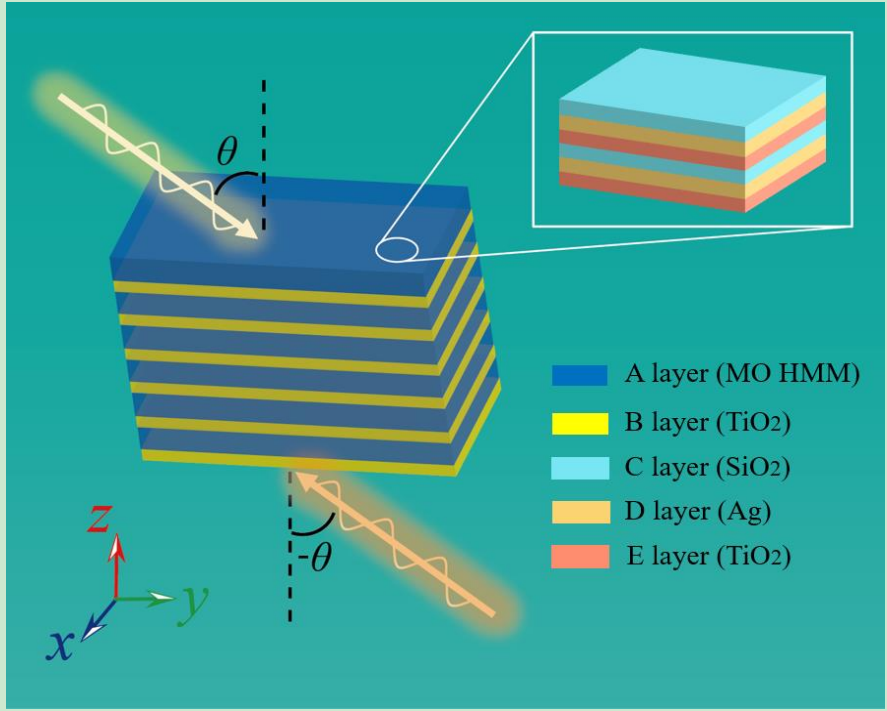

Fig. 1. Schematics of the MO hypercrystal consists of alternating layers of A (MO HMM) and B (dielectric). MO HMM is mimicked by the periodic subwavelength CDE (dielectric/MO metal/dielectric) stacks. The forward and backward incident electromagnetic wave launch into the structure with the incident angle $\theta$ and $-\theta$.

The most remarkable feature of MO PCs is that they have nonreciprocal photonic bandgaps [38, 39]. For the 1D PCs, the Bragg condition of the bandgap is given by

$$\Phi = (k_{Az}d_A + k_{Bz}d_B)|_{\omega_{Brg}} = m\pi, \quad (1)$$

where $\Phi$ denotes the propagating phase in a unit cell and $k_{Az}$ and $k_{Bz}$ represent the $z$ component of the wave vectors within A and B layers, respectively. According to $\partial\Phi/\partial\theta = (\partial\Phi/\partial k_x)(\partial k_x/\partial\theta)$, the position of bandgap can be directly determined by $\partial\Phi/\partial k_x$ when the incident angle $\theta$ is changed, because $\partial k_x/\partial\theta$ is always positive.

Especially, $\partial\Phi/\partial k_x$ is equal to $d_A(\partial k_{Az}/\partial k_x) + d_B(\partial k_{Bz}/\partial k_x)$ and the relation between $k_{jz}$ ($j = A\ or\ B$) and $k_x$ can be determined by the IFC of the media. The dispersion relation of the dielectric in the x-z plane is $k_x^2/\varepsilon_z + k_z^2/\varepsilon_x = k_0^2$, where $k_x = k_0 Sin\theta$ is the parallel wave vector. $k_0 = \omega/c$ is the wave vector in vacuum. The corresponding IFC is a circle, in which the wave vector in the propagating direction $k_z$ will decrease as the incident angle $\theta$ increase. For the traditional all-dielectric PC, both $\partial k_{Az}/\partial k_{Ax}$ and $\partial k_{Bz}/\partial k_{Bx}$ are negative, as shown in Fig. 2(a). Therefore, to maintain the Bragg condition, the bandgap will shift toward a higher frequency as the incidence angle $\theta$ increases. However, for HMM with hyperbolic-type IFC, the wave vector in the propagating direction $k_z$ will increase as the incident angle $\theta$ increase, which is shown in Fig. 2(b). When A layer is replaced by HMM, $\partial k_{Az}/\partial k_{Ax}$ and $\partial k_{Bz}/\partial k_{Bx}$ are positive and negative, respectively. The moving direction of bandgap of the 1D PC containing HMMs depends on the competition between $d_A(\partial k_{Az}/\partial k_x)$ and $d_B(\partial k_{Bz}/\partial k_x)$ [31].

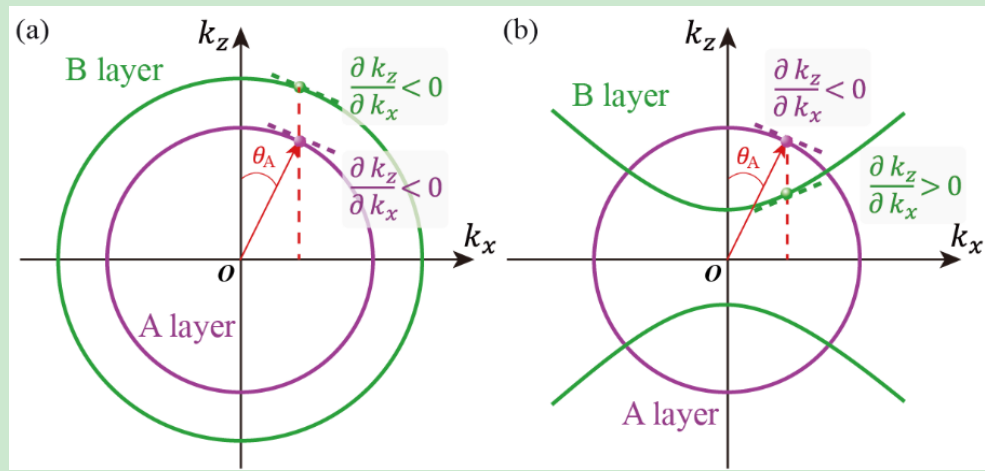

Fig. 2. IFCs of the components A and B layers for tranditional all-dielectric PC (a) and hypercrystal (b), shown as blue and orange lines, respectively. $\partial k_z/\partial k_x$ is positive (negative) for the HMM (dielectric).

Nonreciprocal transmission is very important for electromagnetic wave control [40-43]. It can be used to design key components of modern optical communication systems, such as isolators and circulators [44, 45]. Recently, the significant nonreciprocal transmission based on the enhanced MO effect in HMM has become an active topic of scientific research [46-48]. In this work, we consider the MO hypercrystal which breaks the time-reversal symmetry and study the flexible control of the nonreciprocal bandgap. Supposing that the magnetization under an applied magnetic field is in the $y$ direction, the permittivity tensor $\bar{\varepsilon}_D$ of MO metal (Ag) can be described as [49]:

$$\bar{\varepsilon}_D = \begin{pmatrix} \varepsilon_{xx} & 0 & i\Delta_D \\ 0 & \varepsilon_{yy} & 0 \\ -i\Delta_D & 0 & \varepsilon_{zz} \end{pmatrix}, \quad (2)$$

where $\varepsilon_{xx} = \varepsilon_{zz} = \varepsilon_\infty - \frac{\omega_p^2}{(\omega+i\gamma)^2 - \omega_B^2}\left(1 + i\frac{\gamma}{\omega}\right) = \varepsilon_D$ and $\varepsilon_{yy} = \varepsilon_\infty - \frac{\omega_p^2}{\omega(\omega+i\gamma)}$. $\varepsilon_\infty = 4.09$ is the high-frequency permittivity; $\omega_p = (N_0 e^2/\varepsilon_0 m^*)^{\frac{1}{2}} = 1.33 \times 10^{16}\ rad/s$ presents the bulk plasma frequency, where $N_0$ denotes the free electron density and $m^*$ is the effective electronic mass; $\omega_B = \frac{e}{m^*}B$ is the cyclotron frequency; $\gamma = 1.13 \times 10^{14}\ rad/s$ is the damping frequency; $\Delta_D = i\frac{\omega_B}{\omega}\frac{\omega_p^2}{(\omega+i\gamma)^2 - \omega_B^2}$ means the strength of MO activity [49]. Without the loss of generality, we consider the lossless case with $\gamma = 0$. Supposing that a transverse-magnetic wave ($E_x, E_z, H_y$) propagates in the $x$-$z$ plane. Within an EMT under the condition of long-wave approximation, the MO HMM with subwavelength unit-cells (CDE)$_M$ can be regarded as an effective homogeneous medium, characterized by macroscopic EM parameters [37, 50]:

$$\tilde{\varepsilon}_{Ax} = \varepsilon_{Ax}\left[1 + k_{Ax}d\frac{\Delta_D}{\varepsilon_D}\frac{f_C f_D f_E}{\varepsilon_{Ax}}(\varepsilon_E - \varepsilon_C)\right],$$

$$\tilde{\varepsilon}_{Az} = \varepsilon_{Az}\left[\frac{1 + k_{Ax}d\frac{\Delta_D f_1 f_2 f_3}{\varepsilon_D}(\varepsilon_E - \varepsilon_C)}{1 + k_{Ax}d\frac{\Delta_D f_C f_D f_E}{\varepsilon_D}\varepsilon_{Az}\left(\frac{\varepsilon_E}{\varepsilon_C} - \frac{\varepsilon_C}{\varepsilon_E}\right)}\right], \text{ and } \mu_y = 1 \quad (3)$$

where $\varepsilon_{Ax} = \varepsilon_C f_C + \varepsilon_D f_D + \varepsilon_E f_E$ and $\varepsilon_{Az} = (\varepsilon_C/f_C + \varepsilon_D/f_D + \varepsilon_E/f_E)^{-1}$. Especially, $f_i = d_i/d$, $(i = C, D, E)$ denotes the filling ratio of component layer of MO HMM. The corresponding dispersion relation of the MO HMM can be written as [37, 51]

$$\frac{k_{Ax}^2}{\tilde{\varepsilon}_{Az}} + \frac{k_{Az}^2}{\tilde{\varepsilon}_{Ax}} = k_0^2. \quad (4)$$

Especially, when the wavelength is $\lambda = 335\ nm$, the corresponding IFCs of nonreciprocal HMMs under different magnetizations is shown in Fig. 3(a). The thickness of the unit cell of MO HMM is $d = 50\ nm$. The thickness of the different component layers is $d_{SiO_2} = 0.35d$, $d_{Ag} = 0.41d$, and $d_{TiO_2} = 0.24d$, respectively. When magnetization of the MO metal is not considered ($\Delta_D=0$), the IFC of the effective MO HMM exhibits a typical hyperbolic curve. However, with the increase of the intensity of the external magnetic field, the magnetization of MO metal $\Delta_D$ will increase, which directly affects the IFC of MO HMM. Importantly, the topological transition of dispersion from an open hyperbolic IFC to a closed elliptical IFC is realized for the forward incident electromagnetic waves, as shown in Fig. 3(a). The enlarged IFC of MO HMM for $\Delta_D=0.5\varepsilon_D$ is shown in Fig. 3(b). Especially, $S = \partial k_z/\partial k_x$ is calculated for the forward and backward incident waves, which is a key parameter to realize the control of nonreciprocal bandgap. Red and blue of the curve correspond to $S > 0$ and $S < 0$, respectively. Moreover, based on Eq. (4), we calculate the 3D dispersion relationship of the MO HMM, as shown in Fig. 3(c). It can be clearly seen that the nonreciprocal IFC shown in Fig. 3(a) is preserved in the wide band. Supposing that a TM wave impacting to the MO hypercrystal shown in Fig. 1, the transmission of the structure can be calculated based on the transfer matrix method [52]. The thickness of layer A (MO HMM) and layer B (TiO$_2$) is $d_A$ = 250 nm and $d_B$ = 40 nm, respectively. The permittivity of layer B (TiO$_2$) is $\varepsilon_B$ = 6.55 [53]. From the transmission spectrum in Fig. 3(d), one can see that the bandgap for the forward incident case ($\theta > 0$) is redshifted first within (0º, 45º) and then blueshifted within (45º, 90º) along with the increase of incident angle $\theta$, which is marked by the pink arrows. However, the moving direction of the bandgap for the backward incident case ($\theta < 0$) is opposite. With the increase of incident angle $\theta$, the bandgap is blueshifted within (-30º, 0º) first and then redshifted within (-90º, -30º), which is marked by the cyan arrows.

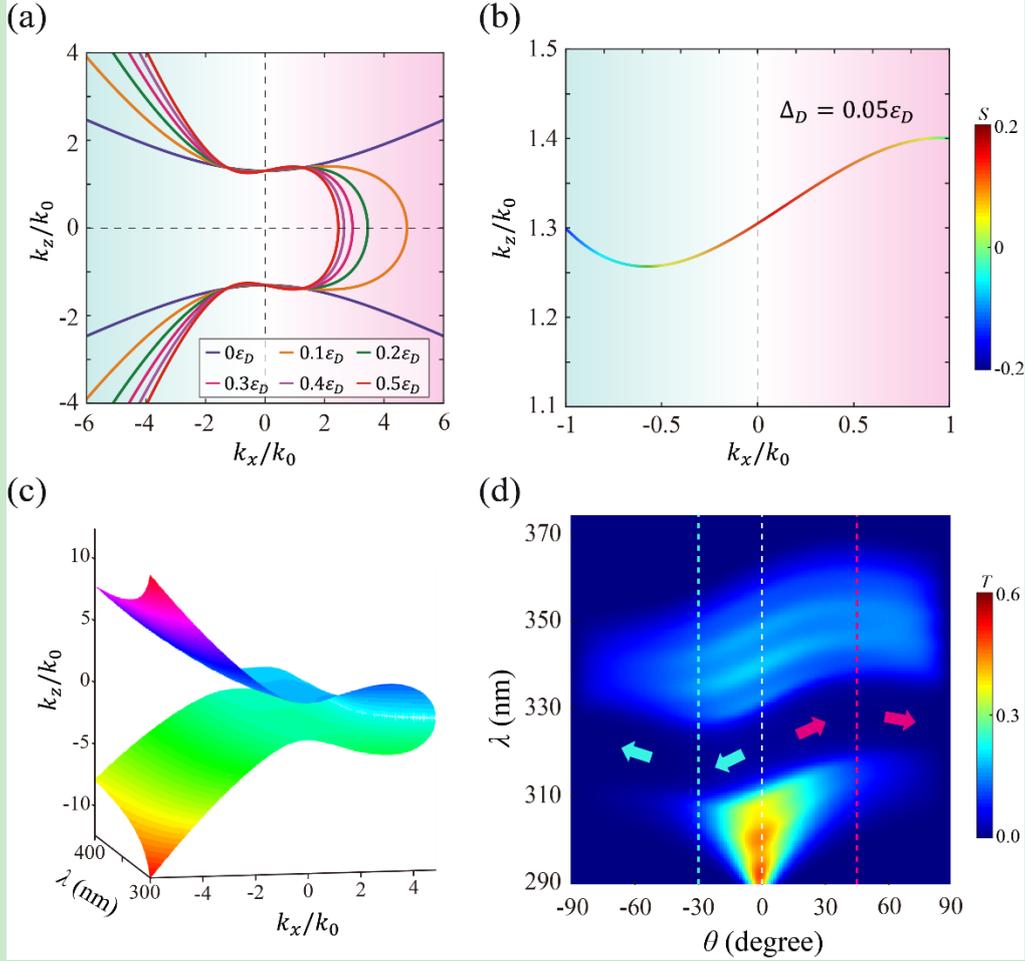

Fig. 3. (a) IFCs of MO HMM (mimicked by a SiO$_2$/Ag/TiO$_2$ multilayer structure) for different magnetization at $\lambda = 335\ nm$. The thickness of the different layers is $d_{SiO_2} = 0.35d$, $d_{Ag} = 0.41d$, and $d_{TiO_2} = 0.24d$, respectively. (b) Amplified IFC of MO HMM for $\Delta_D=0.5\varepsilon_D$. Color denotes the value of $S = \partial k_z/\partial k_x$. (c) 3D dispersion relationships of the- MO HMMs for $\Delta_D=0.5\varepsilon_D$. (d) The transmission spectra of the MO hypercrystal (AB)$_6$ for $\Delta_D=0.5\varepsilon_D$. The thickness of layer A (MO HMM) and layer B (TiO$_2$) is 250 nm and 40 nm, respectively. The moving direction of the bandgap with the increase of the incident angle is indicated by the arrows.

Compared with Figs. 3(b) and 3(d), we can find that the nonmonotonic bandgap of MO hypercrystal can be well predicted by the IFC of MO HMM, which provides a good way to flexibly control the photonic bandgap in photonic engineering. In order to illustrate the flexibility of bandgap control, we further study the monotonic nonreciprocal bandgap realized by MO hypercrystal. For convenience of discussion, we define $u = k_0 d_A \frac{f_1 f_2 f_3}{\varepsilon_{Ax}}(\frac{\varepsilon_E}{\varepsilon_C} - \frac{\varepsilon_C}{\varepsilon_E})$ and $v = k_0 d_A \frac{f_1 f_2 f_3}{\varepsilon_{Ax}}(\varepsilon_E - \varepsilon_C)$. The dispersion relation of MO HMM in Eq. (4) can be written as

$$\frac{k_{Az}^2}{k_0^2 \varepsilon_{Ax}} + \frac{k_{Ax}^2}{k_0^2 \varepsilon_{Az}} + u\frac{\Delta_D}{\varepsilon_D k_0^3} k_{Ax}^3 = 1 + v\frac{\Delta_D}{\varepsilon_D}\frac{k_{Ax}}{k_0}. \tag{5}$$

Considering the critical condition $S_A = \partial k_z / \partial k_x = 0$, we can obtain

$$3u \frac{\Delta_D}{\varepsilon_D} \left(\frac{k_{Ax}}{k_0}\right)^2 + \frac{2k_{Ax}}{k_0 \varepsilon_{Az}} - v \frac{\Delta_D}{\varepsilon_D} = 0. \tag{6}$$

The two corresponding extreme points are $k_{Ax1} = \left[-\frac{1}{\varepsilon_{Az}} + \sqrt{\frac{1}{\varepsilon_{Az}^2} + 3uv\left(\frac{\Delta_D}{\varepsilon_D}\right)^2}\right] / 3u \frac{\Delta_D}{\varepsilon_D}$ and $k_{Ax2} = \left[-\frac{1}{\varepsilon_{Az}} - \sqrt{\frac{1}{\varepsilon_{Az}^2} + 3uv\left(\frac{\Delta_D}{\varepsilon_D}\right)^2}\right] / 3u \frac{\Delta_D}{\varepsilon_D}$. In order to guarantee the monotonicity of the IFC ($S_A > 0$) for MO HMM, the conditions $k_{Ax1} \geq 1$ and $k_{Ax2} \leq 1$ must be satisfied, thus we can obtain the monotonic condition of IFC for MO HMM as

$$\alpha = (v - 3u) \frac{\Delta_D}{\varepsilon_D} - \left|\frac{2}{\varepsilon_{Az}}\right| \geq 0. \tag{7}$$

Figure 4(a) shows three different MO HMMs with the configuration $SiO_2/Ag/TiO_2$ (orange line), $SiO_2/Ag/Si$ (blue line) and $TiO_2/Ag/Si$ (green line), respectively. The permittivity of $SiO_2$ and $Si$ is 2.13 and 12.11 [53]. The thickness of the unit cell of MO HMM is $d = 36\ nm$. The thickness of different component layers is $d_C = 0.35d$, $d_D = 0.41d$, and $d_E = 0.24d$, respectively. Especially, $\alpha$ is positive in a broadband regime of MO HMM based on $TiO_2/Ag/Si$ configuration, and the shaded region in Fig. 4(a) is enlarged for see in Fig. 4(b). $\alpha = 0$ is marked by the dashed line. The effective bandwidth satisfying the monotonic condition in Eq. (7) is from $\lambda = 423$ nm to $\lambda = 472$ nm. Therefore, we can use the $TiO_2/Ag/Si$ stacks to mimic MO HMM in the broadband region, and further study the monotonic nonreciprocal bandgap in the MO hypercrystal.

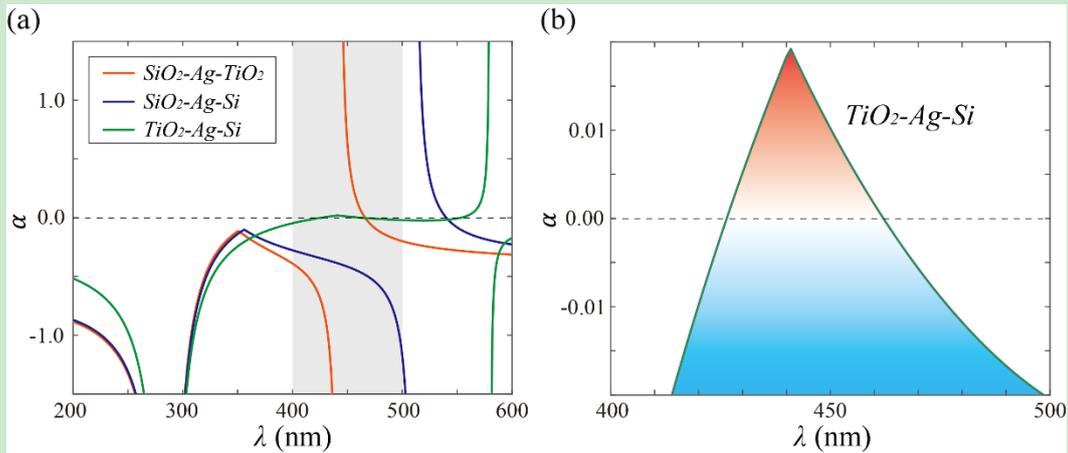

Fig. 4. (a) The $\alpha$ spectrum for three different MO HMMs: $SiO_2/Ag/TiO_2$ (orange line), $SiO_2/Ag/Si$ (blue line) and $TiO_2/Ag/Si$ (green line). The thickness of the different layers is $d_C = 0.35d$, $d_D = 0.41d$, and $d_E = 0.24d$, respectively. (b) Enlarged $\alpha$ spectrum of MO HMM composed of $SiO_2/Ag/TiO_2$ multilayer structure. $\alpha = 0$ is marked by the dashed line.

Taking the wavelength $\lambda = 438\ nm$ for example, the corresponding IFCs of nonreciprocal HMMs (TiO$_2$/Ag/Si configuration) under different magnetizations is shown in Fig. 5(a). When magnetization of the MO metal is not considered ($\Delta_D$=0), the IFC of the effective MO HMM are two flat lines, which can be used to achieve the interesting collimation [53] and long-range energy transfer [28]. Similar to the typical hyperbolic dispersion in Fig. 3(a), the flat IFC of the MO HMM also changes with the increase of the magnetization of MO metal $\Delta_D$. Interestingly, from the enlarged IFC of MO HMM ($\Delta_D$=0.5$\varepsilon_D$) in Fig. 5(b), we can clearly see that $S > 0$ is always positive, which is marked by the color. Especially, the nonreciprocal monotonic property is preserved in the wide band, as shown in Fig. 5(c). Furthermore, we calculate the transmission spectrum of the MO hypercrystal in Fig. 5(d). With the increase of incident angle $\theta$, the bandgap for the forward incident case ($\theta > 0$) is redshifted within (0º, 90º), which is marked by the pink arrow; with the increase of incident angle $\theta$, the bandgap for the backward incident case ($\theta < 0$) is blueshifted within (-90º, 0º). which is marked by the cyan arrow. Therefore, the nonreciprocal monotonic bandgap is demonstrated in the MO hypercrystal.

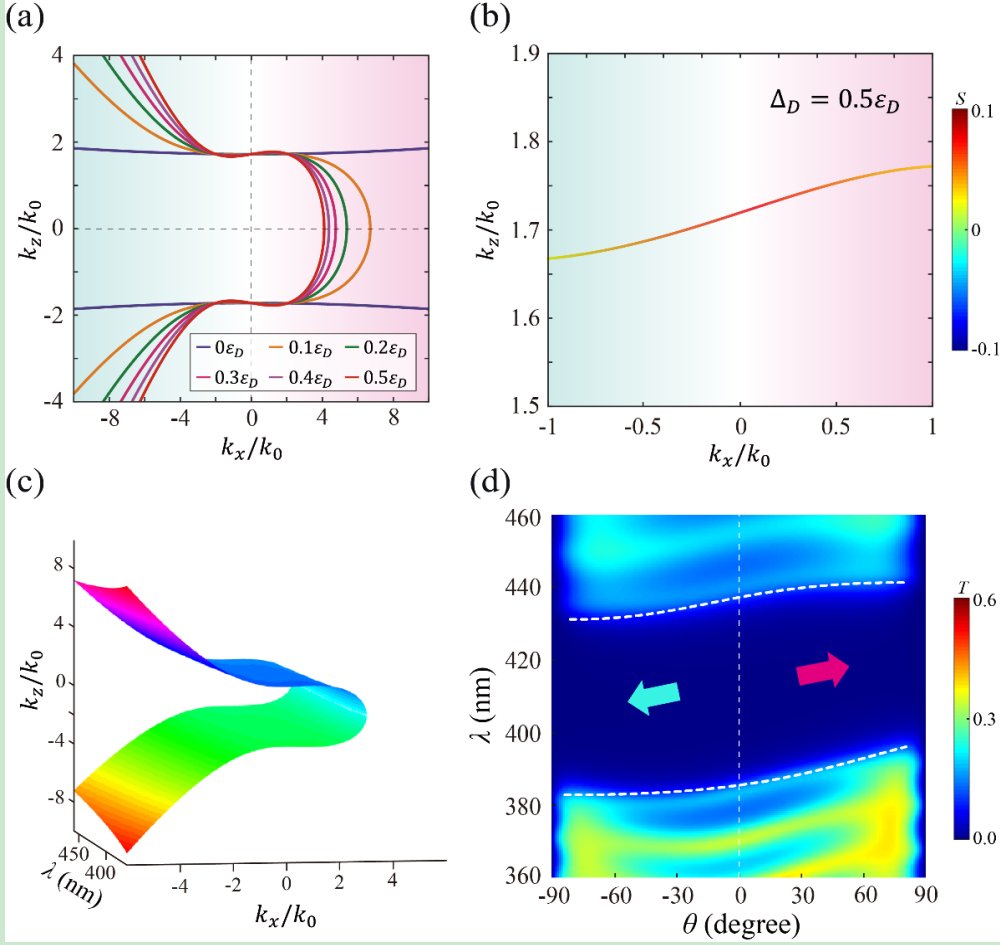

Fig. 5. (a) IFCs of MO HMM (mimicked by a TiO$_2$/Ag/Si multilayer structure) for different magnetization at $\lambda = 438\ nm$. The thickness of the different layers is $d_{TiO_2} = 0.35d$, $d_{Ag} = 0.41d$, and $d_{TiO_2} = 0.24d$, respectively. (b) Amplified IFC of MO HMM for $\Delta_D=0.5\varepsilon_D$. Color denotes the value of $S = \partial k_z/\partial k_x$. (c) 3D dispersion relationships of the MO HMMs for $\Delta_D=0.5\varepsilon_D$. (d) The transmission spectra of the MO hypercrystal (AB)$_6$ for $\Delta_D=0.5\varepsilon_D$. The thickness of layer A (MO HMM) and layer B (TiO$_2$) is 180 nm and 20 nm, respectively. The moving direction of the nonreciprocal bandgap with the increase of the incident angle is indicated by the arrows.

## III. UNIDIRECTIONAL WIDE-ANGLE ABSORBERS BASED ON THE NONRECIPROCAL OMNIDIRECTIONAL CAVITY MODE IN MO HYPERCRYSTAL

Recently, the perfect absorber based on guided resonance of a photonic hypercrystal has been proposed [55]. One of the most important applications of the nonreciprocal bandgap is to realize the unidirectional absorber based on the optical cavity mode. However, it is known that the wavelength of conventional cavity mode in the PC will shift toward short wavelengths with the increase of the incident angle. Therefore, the design of related omnidirectional devices, especially unidirectional and wide-angle optical absorbers, is still a challenge [51, 56]. In this section, we study the nonreciprocal omnidirectional cavity mode in MO hypercrystal. The schematic of the designed MO hypercrystal $(AB)_m F(AB)_{10-m}$ is given in Fig. 6(a), where $m$ denotes the position of the defect layer F in the MO hyperctrystal $(AB)_{10}$. The defect layer F is selected as Au. The permittivity of Au is $\varepsilon_F = \varepsilon_\infty - \frac{\omega_p^2}{\omega^2 + i\gamma_F \omega}$, where $\varepsilon_\infty = 9.1$, $\omega_p = 1.38 \times 10^{16}\ rad/s$, and $\gamma_F = 3.23 \times 10^{13}\ rad/s$ [59, 60]. The MO HMM is realized by the SiO$_2$/Ag/TiO$_2$ configuration. The thickness of the unit cell of MO HMM is $d = 40\ nm$. The thickness of different component layers is $d_{TiO_2} = 0.3d$, $d_{Ag} = 0.53d$, and $d_{Si} = 0.17d$, respectively. We compare the absorption of the cavity mode depending on the position of the defect layer in Fig. 6(b). The magnetization is $\Delta_D = 0.5\varepsilon_D$. The thickness of the different layers in MO HMM is $d_{TiO_2} = 0.3d$, $d_{Ag} = 0.53d$, and $d_{Si} = 0.17d$, respectively. The thickness of layer A (MO HMM), layer B (TiO$_2$) and layer F (Au) is 200 nm, 70 nm and 50 nm, respectively. It can be clearly seen that the system can achieve perfect absorption for different incident angles ($\theta = 0^o$, $30^o$, and $60^o$) when $m$ is 2 and 8. We take $m = 2$ for example, and the MO hypercrystal corresponds to $(AB)_2F(AB)_8$. The corresponding transmission spectrum is shown in Fig, 6(c). One can see that the cavity mode is blueshifted for backward incident case, while the cavity mode nearly remains unchanged with the change of incident angle for forward incident case. The results show high (low) absorption about 0.99 (0.25) in an angle range of 20-75 degrees for the forward (backward) incident light at the wavelength of 367 nm. Therefore, the nonreciprocal omnidirectional cavity mode is demonstrated at $\lambda = 367\ nm$ in the MO hypercrystal, which is shown in Fig. 6(d). Figure 6(e) shows the absorption spectrum of the MO hypercrystal for forward (the red line) and backward (the blue line) propagations. At the working wavelength $\lambda = 367\ nm$, the absorption of MO hypercrystal for forward case with $\theta = 30^o$ and backward cases with $\theta = -30^o$ is 1 and 0.22, respectively. Moreover, the nonreciprocal omnidirectional absorption has also been demonstrated for forward case with $\theta = 60^o$ and backward case with $\theta = -60^o$ in Fig. 6(f).

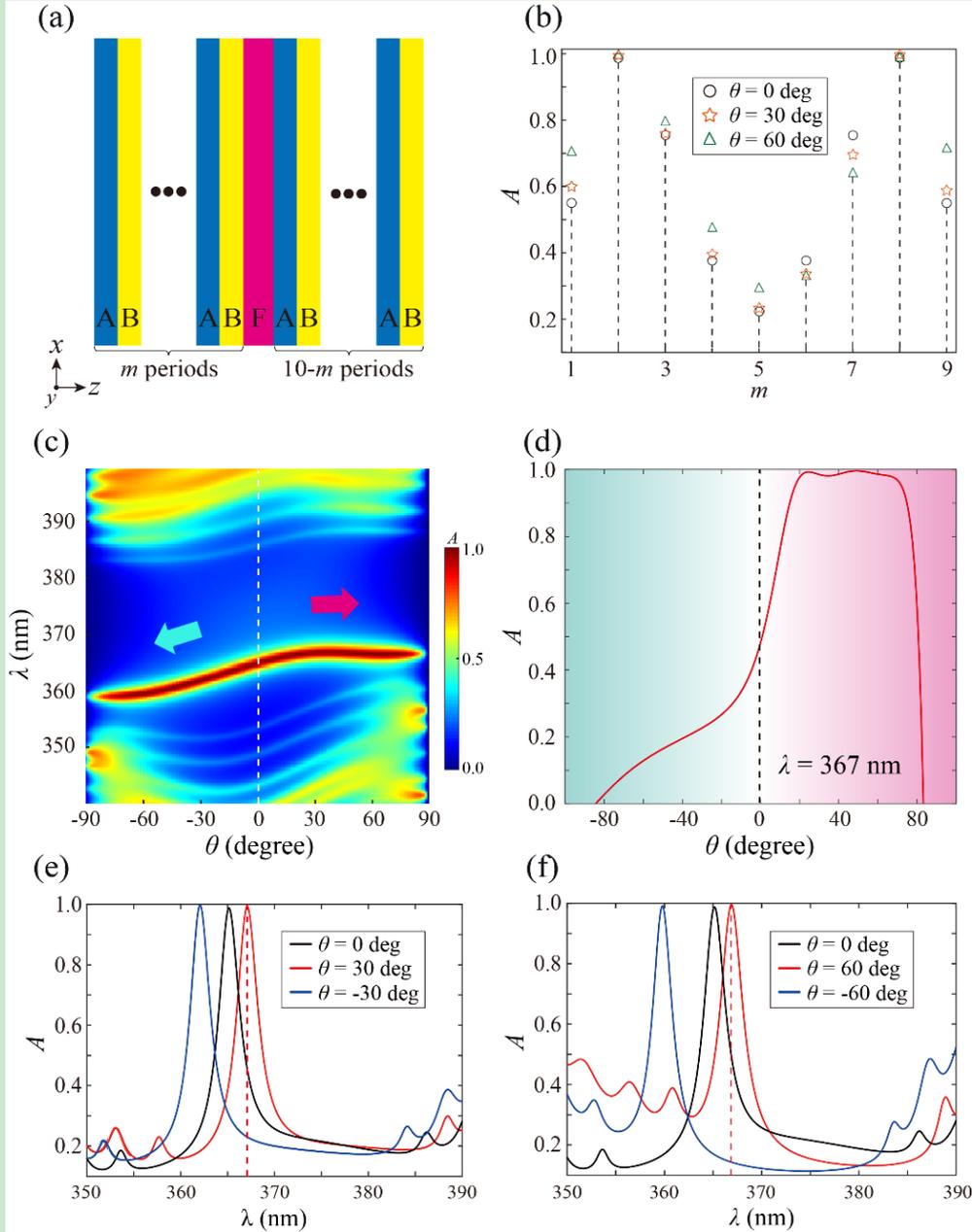

Fig. 6. (a) Schematic of the 1D MO hypercrystal with a defect layer F: $(AB)_mF(AB)_{10-m}$. (b) The absorption of the defect modes for the defect layer placed in different positions of the MO hypercrystal. The magnetization is $\Delta_D=0.5\varepsilon_D$. The thickness of the different layers in MO HMM is $d_{TiO_2}=0.3d$, $d_{Ag}=0.53d$, and $d_{Si}=0.17d$, respectively. The thickness of layer A (MO HMM), layer B (TiO$_2$) and layer F (Au) is 200 nm, 70 nm and 50 nm, respectively. The absorption of the cavity modes for the incident angle $\theta=0^o$, $\theta=30^o$, and $\theta=60^o$ are marked by circles, stars, and triangles. (c) The absorption spectra of the MO hypercrystal with $m=2$. The thickness of layer A (MO HMM) and layer B (TiO$_2$) is 180 nm and 20 nm, respectively. The moving direction of the nonreciprocal cavity mode with the increase of the incident angle is indicated by the arrows. (d) Absorption of the MO hypercrystal at $\lambda=438$ nm as a function of the incident angle. (e) Absorption spectrum of the MO hypercrystal for forward (the red line) and backward (the blue line) propagations with $\theta=30^o$ and $\theta=-30^o$, respectively. (f) Similar to (e), but for the forward (the red line) and backward (the blue line) propagations with $\theta=60^o$ and $\theta=-60^o$, respectively.

## IV. CONCLUSION

In summary, we realize nonreciprocal and flexible photonic bandgaps in 1D hypercrystal with effective MO HMM, including the nonmonotonic and monotonic bandgaps. Moreover, unidirectional wide-angle absorber is designed based on the omnidirectional nonreciprocal cavity mode. This work provides a new type of physical mechanism to design efficient wide-angle nonreciprocal omnidirectional absorber. In particular, the related results may be extended to more flexible active systems under external field control [59-61].

## ACKNOWLEDGMENT

This work was supported by the National Key R&D Program of China (Grant No. 2016YFA0301101), the National Natural Science Foundation of China (NSFC; Grant Nos. 12004284, 11974261, 11774261, 91850206, and 61621001), the Natural Science Foundation of Shanghai (Grant Nos. 18JC1410900 and 18ZR1442900), the China Postdoctoral Science Foundation (Grant Nos. 2019TQ0232 and 2019M661605), and the Shanghai Super Postdoctoral Incentive Program.
Shengyu Hu and Juan Song contribute equally to this work.